# Structural Analysis of Si(OEt)$_4$ Deposits on Au(111)/SiO$_2$ Substrates at Nanometer Scale using Focused Electron Beam Induced Deposition

[1]Po-Shi Yuan, [2]Nigel Mason, [2]Maria Pintea, [3]Csarnovics István, [3]Tamás Fodor

[1]University of Darmstadt, Darmstadt, Germany

[2]University of Kent, School of Physical Sciences, Canterbury, United Kingdom

[3]Laboratory of Materials Science, Institute for Nuclear Research, Debrecen, Hungary

**Abstract**

The focused electron beam induced deposition (FEBID) process was used by employing a Gemini SEM with a beam characteristic of 1keV and 24pA for the deposition of pillars and line shaped deposits with heights between 9nm to 1µm and widths from 5nm to 0.5µm. All structures have been analysed to their composition looking at a desired Si : O : C content of 1: 2 : 0. The C content of the structure was found to be ~over 60% for older deposits kept in air (~at room temperature) and less than 50% for younger deposits, only 12 hours old. Using a deposition of Si(OEt)$_4$ at high rates and a deposition temperature of under 0 degC, an Si content of our structure between 10at% and 15at% (compositional percentage) was obtained.

The FEBID structures have been deposited on Au(111) over an SiO$_2$ wafer. The Au(111) was chosen as a substrate for the deposition of Si(OEt)$_4$ due to its structural and morphological properties, with its surface granulation following a Chevron pattern, and the Au(111) defects having a higher contribution to the change in the composition of the final content of the structure with the increase in O ratio and a reduction in the shapes heights.

**Introduction**

The focused electron beam induced deposition (FEBID) was first discovered in 1962 when molecular deposits in the mass spectrometer equipment result of the electron beam induced chemistry on the molecules of the compound have been found. Since 1962, the technique has received more attention, in the past 10-15 years progressing to being a viable manufacturing method of circuits, nanomaterials, semiconductors and with implications in a wide range of medical applications. Principles of FEBID techniques have been researched by scientists in *Huth et al (2018) [6], Toth et al (2015) [7], Randolph et al (2006) [8], Randolph et al (2005) [36], Huth et al (2012) [37], Thorman et al (2015) [38], De Teresa et al (2016) [55]*, providing information on nanostructure deposits of Fe(CO)$_5$ and Co$_2$(CO)$_8$ with very high deposition purities of the structures with values of ~98at% (Fe(CO)$_5$) *[37]* and 95at% (Co$_2$(CO)$_8$) *[55]*. High purity structures of bimetallic compounds with values in the range of ~80at% have been obtained by *Kumar et al (2018) [56]* irradiating deposited layers of HFeCo$_3$(CO)$_{12}$ and exposing them to room temperature where a desorption of the remaining ligands to CO and H from the resulting HFeCo$_3$(CO)$_3$ was observed. The FEBID method is mostly used with SEM deposition of nanostructures and in-situ XPS, AFM and EDX studies of the deposited nanostructures for DUV/EUVL mask repair or complex - 3D nanostructure depositions, materials ranging from carbonyls, acetates, acetylenes, bromides, chlorides and iodises to combined bimetallic or trimetallic precursors.

The silica precursors used for substrate deposition ($SiO_2$ deposition) range from TEOS to $SiCH_x$, $SiNH_x + O_2$ depending on the deposition process. The most common processes used in the past are CVD and ALD, with a very wide pool of applications to radiation in cancer research and radiation therapy (GSH-responsive mesoporous silica nanoparticles *[54]*), new materials development for catalysis and hydrogenation (Pd-containing hydrogenation nanocrystals immobilized in silica precursors *[50]*, $SiO_2$ – $CeO_2$ nanoparticles with heat specific tolerance *[52]*), fibre-optics in telecommunications (periodic mesoporous photoluminescent nanocrystal silicon-silica composites *[53]*) or radiation containment of water, energy generation and uranium storage (SG - TTA + $SiO_2$ with a 98% sorption of uranium(VI) *[49]*). The focused electron beam induced process (FEBIP) of depositing $SiO_2$ from TEOS was optimized using CASINO simulations of the beam parameters *[58], [59]* and gas-phase studies of the $Si(OEt)_4$ fragmentation pathways, earlier EBID studies presenting facts on the deposition of $Si(OEt)_4$ at multiple temperatures as a standalone and mixed with $H_2O$ *[57]* with some success.

**Experimental Section**

**SEM and measurement equipment.** The scanning electron microscope (SEM) used for deposition of the nanostructures, a SEM C1 with a LEO 1500 series Gemini column has a performance of the Leo 1500 series is 1.0nm @ 20kV, WD=2mm to 5.0nm @ 0.2kV, WD=2mm in high vacuum and 1.2nm @ 20kV, 7.0nm @ 1kV in variable pressure mode, with an acceleration voltage in the range 0.1kV to 30kV. The electron source, a Shottky field emitter, is characterized by high beam brightness and low beam energy spread. The SEM is provided with an infrared CCD camera with a focusing distance of 1mm to 50mm and eight pole electromagnetic stigmator, and a stage, a 5 axis eucentric stage with motorized movements on x, y in the range of 75mm and z in the range of 55mm.

The nanostructures analysis was carried out using EDX measurements. The EDX measurements are characterized by an analysis beam voltage of 5keV. The principle behind the EDX functioning is the process of collision between a molecule or an atom with an electron, with the result of the appearance of a hole in the inner shell, while an electron from the outer shell will take its place emitting a set of characteristic X-rays specific to each element from the periodic table and with a characteristic acceleration voltage specific to it, defined by the relation $z_m = 0.033 (E_0^{1.7} – E_c^{1.7}) A/\rho Z$. The Kα value for our $Si(OEt)_4$ (characteristic X-ray) are Si(1.739), O(0.525) and C(0.277) resulting a minimum accelerating voltage of the TEM beam with a value of 5kV.

**XPS.** X-ray photoelectron spectra were obtained using an Al/Mg twin anode non-monochromated radiation source and a Phoibos100 MCD-5 series hemispherical energy analyser produced by SPECS (Berlin). The measurements were conducted with Al K-α (E = 1486eV) rays. The sample was examined as received, mounted directly onto the XPS sample holder. The spectra were processed with CasaXPS (http://www.casaxps.com).

**Computational details.** For the electron trajectory simulations, the CASINO software version v2.42 and v3.4 were used with focus on beam characteristics of the secondary and backscattered electrons involved in the deposition process of the nanostructures.

**Si(OEt)$_4$ Deposition.** The Si(OEt)$_4$ was deposited with different dwell times and loops numbers. A set of two deposits on the Au(111) surface were done, first set was kept 8 months in the air after deposition and a second set was deposited that was analysed after up to 12 hours from the deposition time.

The first set of structures (**Fig 1**) was deposited using the beam characteristics of 24pA beam current and 1keV beam voltage. The parameters of the deposition process are presented in **Table 1**. The set of measurements contains a number of 2 sets of 6 points and 6 sets of 6 lines, done with variable dwell time constants (presented in **Table 1**) at 0.7μs, 0.45μs, 0.35μs and 5μs respectively.

| Deposits Type | Dwell Time (μs) | Loops (no) | Beam Current (pA) | Beam Voltage (eV) |
|---|---|---|---|---|
| 6 points | 2.350/ 12.35/ 22.35/ 32.35/ 42.35/ 52.35 | 1k/ 1.3k/ 2k/ 2.5k/ 3k/ 1.8k/ 1.5k/ 4k | 24 | 1000 |
| 6 points | 10.325/ 22.325/ 24.325/ 26.325/ 30.325 | 1300 | 24 | 1000 |
| 6 lines | 5 | 1300 | 25 | 1000 |
| 6 lines | 0.7 | 500 | 24 | 1000 |
| 6 lines | 0.45 | 500 | 24 | 1000 |
| 6 lines | 0.7 | 700 | 24 | 1000 |
| 6 lines | 0.35 | 1300 | 24 | 1000 |
| 6 lines | 0.7 | 1300 | 24 | 1000 |

Table 1. Deposition of Si(OEt)$_4$ on Au(111)/silica, 8 months old deposits

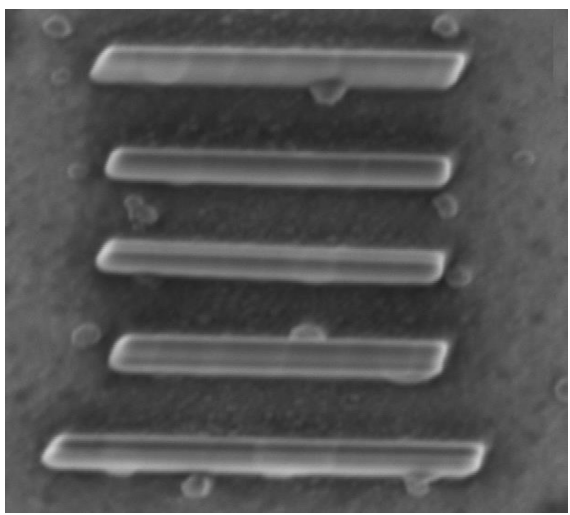

Fig 1. 8 months old deposits of Si(OEt)$_4$ on irregular Au surface, 6 lines (0.7μm; 700μs) top view

The second deposition (**Fig 2**) was done using the beam characteristics of 28pA beam current and 1keV beam voltage. The deposition parameters of the second deposition are presented in **Table 2**. The Au(111)/silica substrate used for the second deposition has different features than the characteristics of the first substrate; it is more homogenous, and the Au(111) surface deposited on silica has a constant layer thickness of 100nm. The first substrate in comparison has a layer thickness of 200nm, is older than the second substrate, with a higher degree of contamination of the surface, dust particles, $NO_2$, H, $H_2O$, grease, irregularities and defects. At defect sites, the silica wafer can be seen through the previous depletion of Au(111), or the layer of Au(111) is reduced in dimension and the signals from the $SiO_2$ wafer can be obtained. The second deposition contains a number of 2 sets of 7 line deposits at 0.7μs and 0.35μs dwell times and 2 sets with 6 and 7 point deposits with a variable dwell time constant of t + 0.2μs for the set of 6 points and t + 1μs for the set of 7 points.

| Deposits Type | Dwell Time (us) | Loops (no) | Beam Current (pA) | Beam Voltage (eV) |
|---|---|---|---|---|
| 7 lines | 0.7 | 650 | 28 | 1000 |
| 7 lines | 0.35 | 1300 | 28 | 1000 |
| 6 points | 4.03/ 4.23/ 4.43/ 4.63/ 4.83/ 5.03 | 1000 | 28 | 1000 |
| 7 points | 0.25/ 1.25/ 2.25/ 3.25/ 4.25/ 5.25/ 6.25 | 2600 | 28 | 1000 |

Table 2. Deposition of Si(OEt)$_4$ on Au(111)/silica, 12 hours old deposits

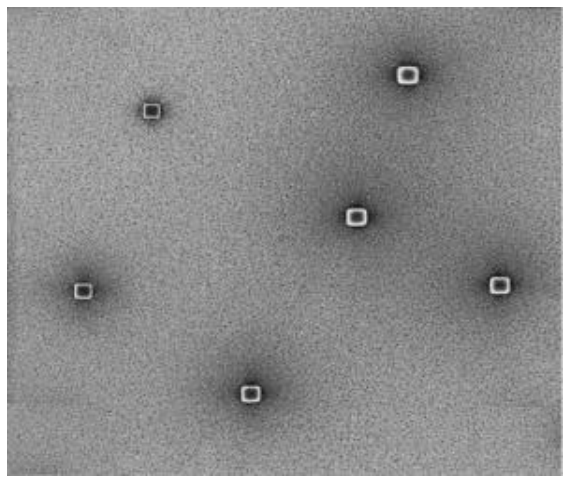

Fig 2. 12 hours old deposits of Si(OEt)$_4$, 6 points top view

The deposits were done at a vacuum chamber pressure of 8.2 x 10$^{-7}$mBar, with a deposition pressure of 1.5 x 10$^{-6}$mBar and a temperature of the Si(OEt)$_4$ precursor in the gas line storage vial of -11°C for the second set of deposits and -25°C for the first set of deposits.

**Results and Discussion**

**Au(111)/Silica Substrate Characterization and Au(111) surface reconstruction**

**Au(111)/Silica Substrate XPS Characterization.** X-ray photoelectron spectroscopy was used for post-deposition substrate characterization, as the instrument setup only allows photoelectron collection from a macroscopic (7 × 20mm) surface area, which is several orders larger than that of the Si-bearing deposits. The survey spectrum shows that the analysed surface consists of gold and some contaminants (see **Fig XP1** and **Table XT1**).

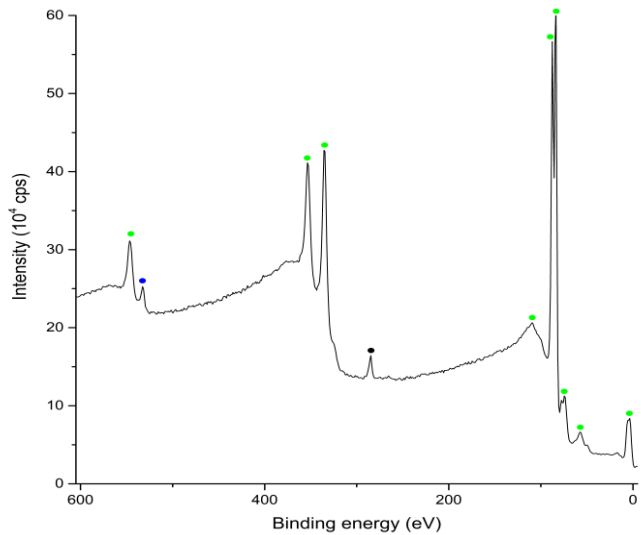

Fig XP1. X-ray photoelectron spectrum of the sample. Photoelectron peak assignment:

black – carbon, blue – oxygen, green – gold

| Element | Oxygen | Carbon | Gold |
|---|---|---|---|
| Concentration | 9.46% | 33.45% | 57.09% |

Table XT1. Elemental composition of the sample surface, as determined by XPS

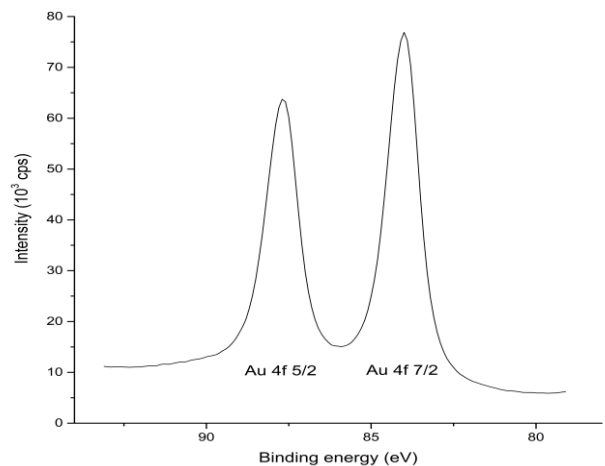

Fig XP2. X-ray photoelectron spectrum of the Au 4f region

The appearance of adsorbed oxygen and carbon are to be expected, as the sample was in prolonged contact with air before the measurement took place and this technique yields information from the outermost 5-10 nanometres of the surface. No signals corresponding to silicon were detected via XPS. This was expected, as the peak intensity is a function of surface coverage and the deposits are nanoscale deposits, constituting less than 1ppm of the total examined surface atoms. The gold substrate itself is composed of the pure element, to the extent that the Au 4f 7/2 peak appears precisely at the literature binding energy value (84.0 eV) and was used for calibration (see **Fig XP2**).

The 200nm (8 months old substrate) and 100nm (12 hours/12 months old substrate) present no signal coming from the $SiO_2$ wafer. Simulations of Au(111) surface *[42], [43]* reveal the presence of (111), (110), (100) and (211) facets of the substrate. While the Au – Au has a bond value of ~2.10 – 3.10Å and a unit cell length of 4.065 x 4.065 x 4.065Å, the underlying SiO$_2$ wafer has the Si – O bond distances with a value of 1.58 - 160Å and the unit cell of 7.16 x 7.16 x 7.16Å, influencing only the organization of the first Au-monolayer on the wafer, forcing the subsequent layers of atoms in a bcc plane configuration. *Hanke and Björk (2013) [44]* do a reconstruction of the Au(111) substrate for a six-layer slab using a 22x√3 lattice with a number of 23 atoms fitted in 22 sites for the fcc and hcp organization of the atoms; imposing the rule that a minimum of six-layers of Au atoms is needed for a full convergence. The organization of the atoms on the SiO$_2$ wafer are following the bcc orientation on the first layer followed by combined fcc and hcp sites, while the last layers are organized in fcc configuration. Earlier studies *[42], [43]* were not able to solve the presence of both hcp and fcc sites as well as the atomistic and electronic degree of freedom important in determining the reactivity of sites and the catalytic activity of the reactions at the surface in the FEBID of the nanostructures. The Au(111) substrates over time present a highly grained surface with visible Au atoms (**Fig XP3**; 8 months old substrate), a reorganization of the surface layers pushing at sites Au atoms with 0.01Å to 0.5Å higher, as well as the process of integration in the lattice of gas atoms or other atoms from the deposited precursor.

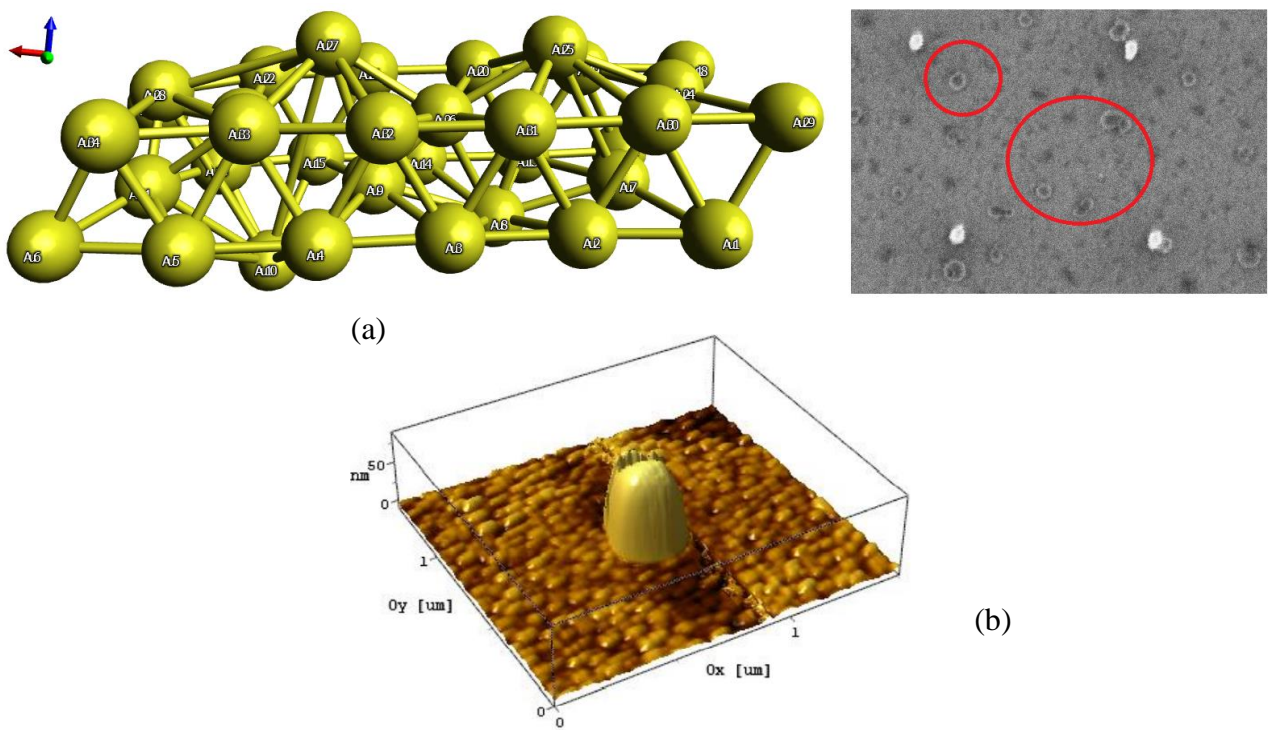

(a)

(b)

Fig XP3. 2-layer Au - substrate optimization view with Au27, Au25, Au10 and Au8 sites higher from the axis line (6 x 3 x 2 atoms) (a); 91nm high pillar shape nanostructure with apparent visible Au-grain surface (b)

**Deposits Analysis.** Si(OEt)$_4$ or under his most common denomination TEOS is one of the most common deposition precursors used in FEBID and SEM assisted processes, CVD, ALD and ALD-CVD for structure deposition at the nanometre scale and for mask repair industry. With a high oxygen content, the Si(OEt)$_4$ is a great candidate to be the chosen precursor for thin layer deposition of SiO$_2$. Safe, with a low content of acids as results of decomposition, non-explosive and not harmful if inhaled in small quantities, although used in enclosed recipients, makes one of the most desired chemical compounds for deposition purposes.

Different deposition processes have been used to create thin layers and well-defined structures on surfaces, such as CVD, first used in 1961 for the deposition of TEOS, LPCVD, APCVD or PECVD [1], all differing in the temperature of deposition of the compound and the use of $O_2$ or $O_3$. During a CVD process of TEOS, the deposition temperature reaches 750°C, for LPCVD process the temperature is reduced to 600°C, while PECVD with the addition of $O_2$ has a nominal temperature of 200°C releasing and removing CO and $CO_2$ and obtaining a structure with a higher resistivity of ~$10^{16}$Ωcm. APCVD compared to PECVD or LPCVD, has the advantage of addition of $O_3$ to the general deposition process and a high reduction in temperature close to ~300°C; the $O_3$ traps the TEOS molecules on the surface depositing higher efficiency thin films and structures with lower contamination and lower stress levels. When using alkoxysilanes as precursors for ALD deposition, the addition of water and the chemisorption of $SiH_4$ on a hydrogenated oxide surface is necessary to break the Si – OR bonds by reaction with a hydroxy – OH group, but by using a - $NH_2$ compound, $H_2N(CH_2)_3Si(OEt)_3$ [2], the deposition of $SiO_2$ can take place without the use of a catalyst. Other sources of $SiO_2$ used in deposition and for analysis studies are $SiH_3$ [3], [4], $SiH_4$ [5], diethylsilane ($Et_2SiH_2$) [6], 1,4 - dislabutane (DBS, $H_3SiCH_2CH_2SiH_3$) [6], 2,4,6,8 - tetramethylcyclotetrasiloxane (TMCTS – R = $CH_3$) [6] and 2,4,6,8 - tetraethylcyclotetrasiloxane (TMCTS – R = $C_2H_5$) [6]. In the focused electron beam induced deposition process (FEBID) [7], [8] with CVD, ALD, PECVD, or by itself, the necessity of adding $O_2$, $O_3$ or any of the hydroxy – OH groups is removed, the use of the electron beam for the fragmentation and breaking of the bonds to form high purity deposits of $SiO_2$ is proven to be a method with high efficiency and purity content of the deposits of 90at%.

**Beam currents and deposition rates.** Two types of deposits were done for the first deposition set of $Si(OEt)_4$ 8 months old structures, line profiles and dot profiles (pillars). The line profiles have heights/widths of 103.2nm – 401nm/280nm – 520nm, 146.2nm – 290.7nm/239nm - 361nm, 283.9 – 311.6nm/330nm – 410nm and 278.6nm – 291.9nm/285 - 309nm for dwell times (μs) of 700/350/450/1000μs. At a first look, the line profiles (**Fig 1**) are better preserved than the dot profiles (**Fig 2**). The AFM measurement and processed images of the deposits show a merging of the dot profiles (b) and an evolution of their structure from a homogenous structural morphology to similar structures though with a higher carbon content and tilting/collapse created in a first instance by the drying of the product and after by the accumulated moisture from the air in conjunction with a much smaller base width than the height value. Similar behaviour was observed in past experimental work by *Randolph et al (2005)* in *[9]*. The values reported by *Randolph and co-workers [9]* are for structures with similar dimensions, deposited at the electron beam energy of 1keV. The orientation of the Au(111) grains of the substrate can have a high influence on the shape and orientation of the resulting structures on the deposition surface. The deposition rate as well as the height of the dot profiles have a big influence on the induced damage on the structure by the N and $H_2O$ in the air, higher height could mean the breakage and collapse of the structure with gravity is due to moisture and not the effect of oxidation on the structure. The width of the deposits is directly proportional to the density of backscattered and secondary electrons emitted from the 1keV electron beam used for deposition. According to that, the halo around the structure and the Si found in the background spectrum would give an indication of how much is deposited near the structure during the structure growth. Values of 0.63nm average have been found for the Au(111) substrate, while the halo around the structures has a thickness value of 0.015nm; the two smaller structures (**Fig 1**) around the main structures are substrate grains covered by thin layers of precursor.

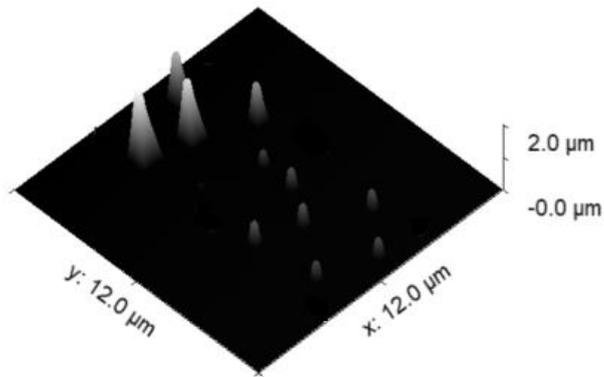

Fig 3. Si(OEt)$_4$ 12 hours old deposits 3D view

The 12 hours old deposits are characterized by no morphologic modifications due to the accumulation of N$_2$ and H$_2$O on the surface, and to less exposure to air and atmospheric pressure, as well as a smoother surface characterized by less defects and kinks where atoms and parts of the fragments agglomerate.

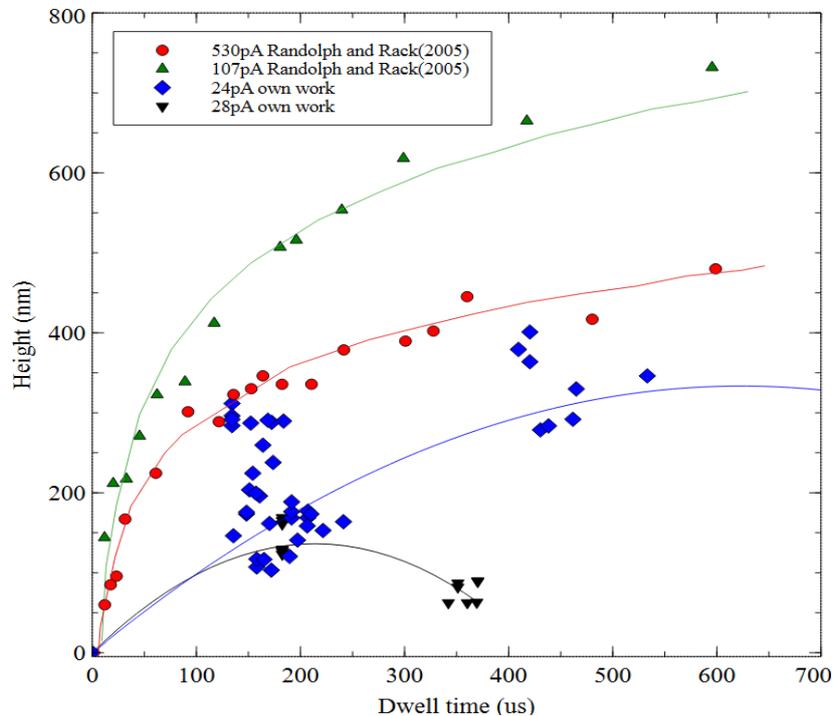

Fig 4. Deposition rates of Si(OEt)$_4$ for different beam current values

The dimensions of the 12 hours old deposits are 62.4 – 89.9nm/310 - 341nm height and 123.1 – 168.8nm/350nm width for a dwell time (μs) of 700us/350μs for the line profiles and 487 - 1628nm height and 248 - 783nm width for a variable dwell time (μs) of the dot profiles in the range 250 - 5030μs, for a beam characteristic of 1keV. The 12 hours old deposition and background analysis shows a 10nm substrate to the structure growth compared to the older substrate from the 1st deposition with a value over 630nm, though around the structures in the 2nd deposition no extra material can be found. *Randolph, Fowlkes and Rack (2005) [9]* declared higher deposition cross-sections values than obtained from our experimental work, almost double in height for currents of 107pA and 530pA (**Fig 4**). The deposition times (μs) of the structures and their widths (nm) and heights (nm) are presented in the additional file supporting the article.

**Casino simulations of beam characteristics.** Further surface studies involving CASINO simulations (**Fig 5**) of electron distributions show a maximum radius of visible electrons around the structure of up to 9nm with the highest distribution between 1nm and 2nm. Backscattered electrons and secondary electrons have energies as high as 200eV, higher than the low electron DEA, that could break ligands and form additional negative and positive anions that would deposit secondary structures around the main structure or create additional contamination on the substrate, e.g., layers of ethyl and methyl, adding to the contamination of the main structures. The simulations at 1keV have lower backscattered radius than the simulations done at 2keV. At 5keV a backscattered radius of 26nm is observed, definitely higher but as well lower in the number of electrons that can create the structure over 1 nm$^2$.

The CASINO simulations have been run on a predefined Au(111) substrate with the 100nm x 100nm x 20nm and a pyramid set to intercept the box with 50nm x 50nm x 50nm at angles (70, 90, 70, 90) with Si(OEt)$_4$ composition. The distribution of the backscattered electrons and secondary electrons is obtained from the backscattered radius, presenting the length of the density of backscattered electrons around the predefined (0, 0, 0) point of the main beam. The CASINO version used was CASINO v.3.3, with comparison to the CASINO v2.42 for planar area surface distribution (cross-section of adsorbed energy of backscattered SEs).

5keV

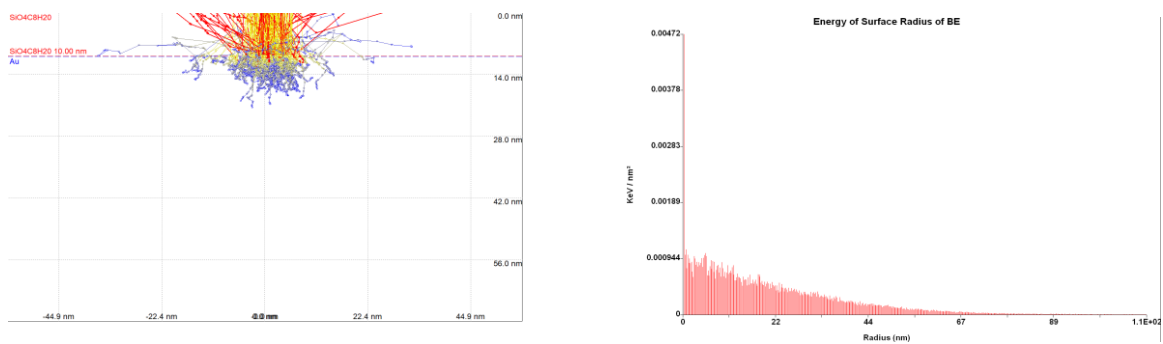

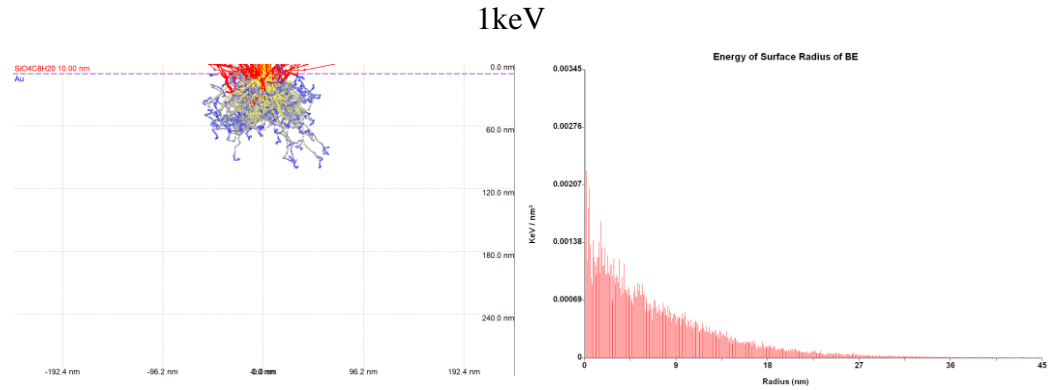

Fig 5. Backscattered electron distribution on surface at 1keV and 5keV

| Beam Voltage (keV) | Maximum energy of surface radius (hits/nm$^{2)}$) | Backscattered Radius (nm) | Energy of surface radius of BSE keV/nm$^2$ |
|---|---|---|---|
| 1 | 0.151 | 1.2 | 0.121 |
| 2 | 0.05316 | 12.2 | 0.08172 |
| 5 | 0.03617 | 26 | 0.134 |

Table 3. Results of electron trajectory simulations using CASINO software for 1keV, 2keV and 5keV

Based on the Monte-Carlo routine of electron-trajectory calculations, the software presents a number of backscattered radii, secondary electron radius, maximum scattered radius, energy of the backscattered electrons and secondary electrons. Earlier studies of backscattered and secondary electron processes in Mott insulators and cathodoluminescence have been run using CASINO v2.42 simulations [34], [35], while newer studies focus on the sensitivity of measurement (2020) [36] and 3D applications in CMOS nanotechnology (2011) [37] using CASINO v3.3. In the present study we intend to obtain simulation data of secondary and backscattered electrons for FEBID nanolithography of Si(OEt)$_4$ precursor used in the deposition of SiO$_2$ nanostructures. The software focuses its algorithm on Markov chains [46] and Voronoi triangulation [47] and uses the splitting of the nanostructure + substrate in 3D triangles (meshing) model developed by Akenine – Möller in 1997 and improved by the addition of all the triangles into a 3D partition tree by Mark de Berg in 2008 [37]. All the structures have defined an inner shape and an outer shape making possible the declaration of different compositions at the top thin layer (oxidation, substrate – nanostructure interactions, thin layer effects).

We observed different distributions of the BSEs on the structure/Au(111) with the change in the PEs energy from 2keV and 5keV (see **Table 3**) with almost 50% in both cases, while for the 1keV the radius is 10 times lower, though the energy of the BSEs on the surface is higher than for the 2keV simulation. Lower energy is observed in the case of the 5keV simulations of 0.134keV/nm$^2$ close in value to the energy obtained for the 1keV beam of 1.121keV/nm$^2$. Simulations of higher energies up to 10keV have been done with CASINO v2.42 in thin Si(OEt)$_4$ layers of no more than 10nm with the energy of the surface radius obtained of 0.00253keV/nm$^2$ and most of the BSEs falling between 8 – 10keV.

To obtain the energy distribution of backscattered electrons, the sample was declared as a 10nm layer with SiO$_4$C$_8$H$_{20}$ composition deposited on an Au(111) substrate using CASINO v2.42, for a number of 200 displayed trajectories. The total and partial cross sections for the electron distributions used the model of Drouin and Gauvin (1993) and ionization potential from the Joy and Luo model (1989).

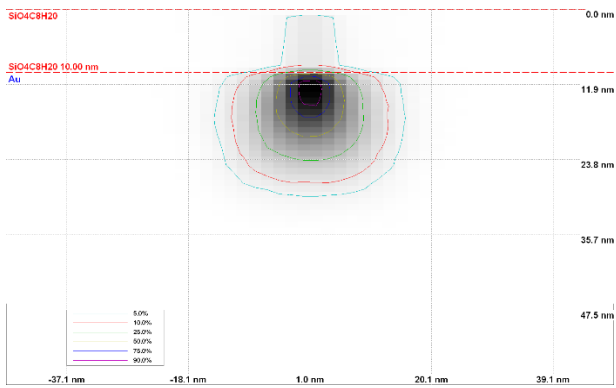
1keV, 1nm

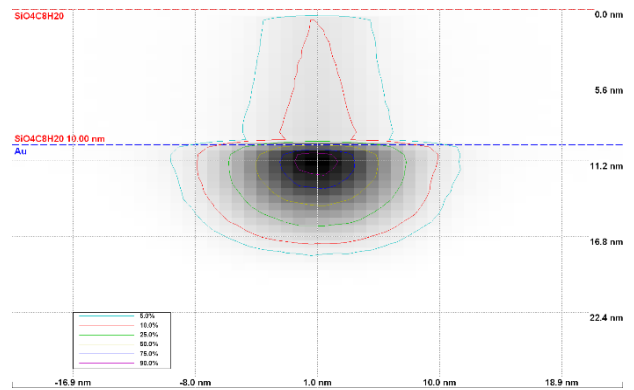
2keV, 1nm

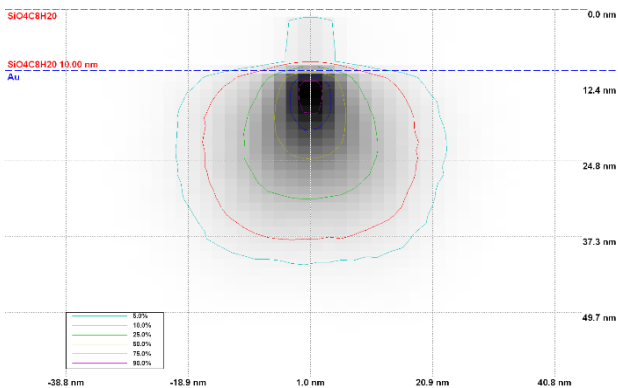
3keV, 1nm

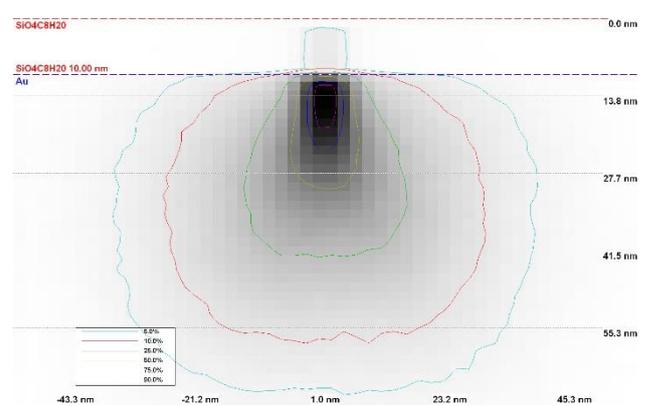
5keV, 1nm

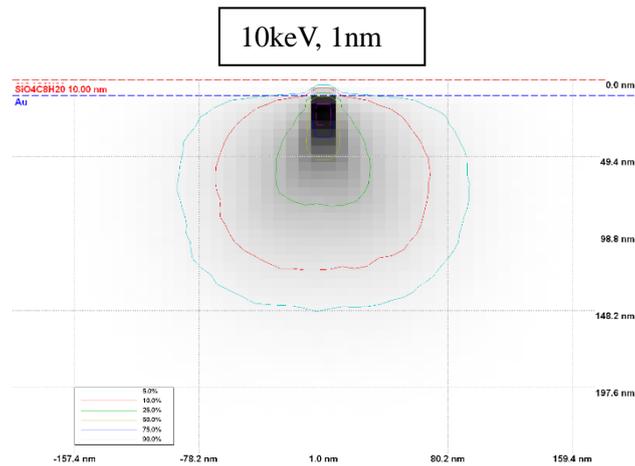

Fig 6. Simulated absorbed energy of the sample and substrate at 1 – 10keV and 1nm from the center of the sample

The simulated absorbed energy is presented in **Fig 6**, it can easily be observed that for 1keV, 3keV and 5keV the presence of 10% distribution lines is lower in the nanostructure compared to 2keV and 10keV where the 10% and 5% lines go upper in the nanostructure with sharper peaks, though for 1keV and 10keV we have contribution from the 25% lines.

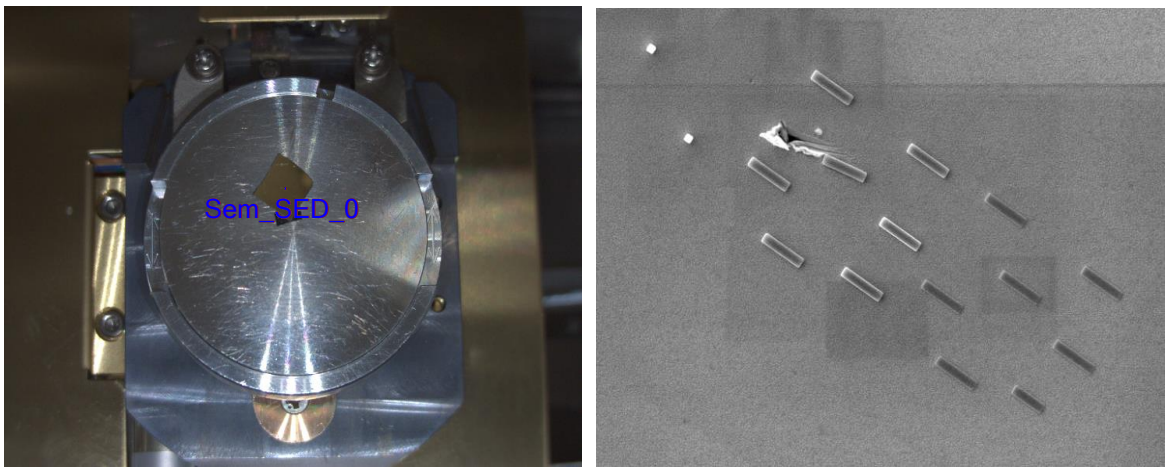

Fig 7. Si(OEt)$_4$ on Au(111)/SiO$_2$ 12 months old sample (left) and top view of the deposits (right)

**Deposits Analysis.** A second set of measurements was carried on the 12 hours old deposits after a period of 12 months (**Fig 7**). The deposits suffered modification over time due to storage and handling, as well as due to the aging of the substrate, the packing of the Au atoms modifies, creating holes and kinks in the structure and undergoing in particular areas phenomena as tasation (settlement), integration and reorientation *[42], [43]*. Some structures are missing due to storage and handling issues data not being available for comparison. The 12 months old measurements have been carried in Laboratory of Materials Science, Institute for Nuclear Research, Debrecen, Hungary, the substrate being shipped there in ambient atmosphere, with no extra handling measures applied. We do not assign the defects on the substrate to the transport or handling, but to the modifications and reorganization of the Au atoms under ambient atmosphere and room temperature. Some of the structures collapsed as a result of prolonged exposure to air. The deposits have been measured at 10deg and 15deg stage tilt to obtain the compositional content of the structures. The heights of the structures were obtained with a certain degree of error, as the EDX measurement is not intended for verification of the nano-deposits height. A separate AFM study to determine the height of the structures was run, ranging in values from 70nm (the smallest) in the shape of a line to 1.684µm (the largest) in the shape of a pillar, and a very large base, almost equal in dimension with the height, with a value of ~1.2µm.

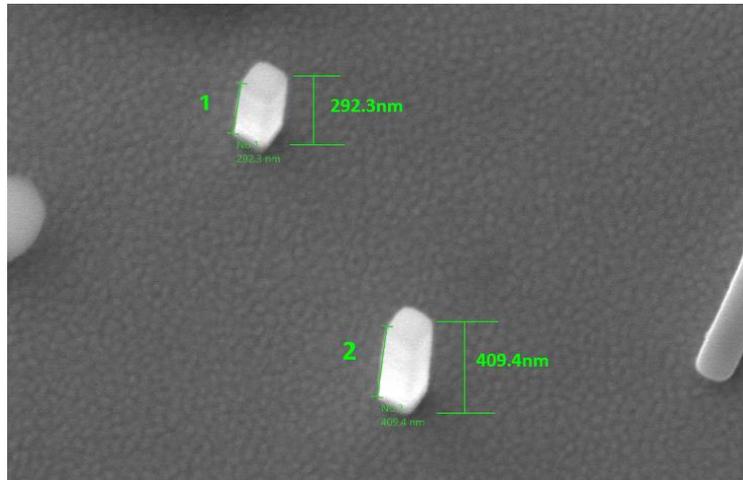

Fig 8. Tilted view to 15deg of the $Si(OEt)_4$ pillar shape structures of 12 months old deposits

For the structure 2 in **Fig 8** an EDX compositional analysis was run (see **Fig 9**) with resulting C content up to 21wt%, 26.11wt% in O and 17.91wt% in Si, with the atomic ratios of 41.73at% C, 38.86at% O and 15.19at% Si. The presence of Au is due to the thin layer of deposit, EDX acquiring data up to 100nm on the substrate.

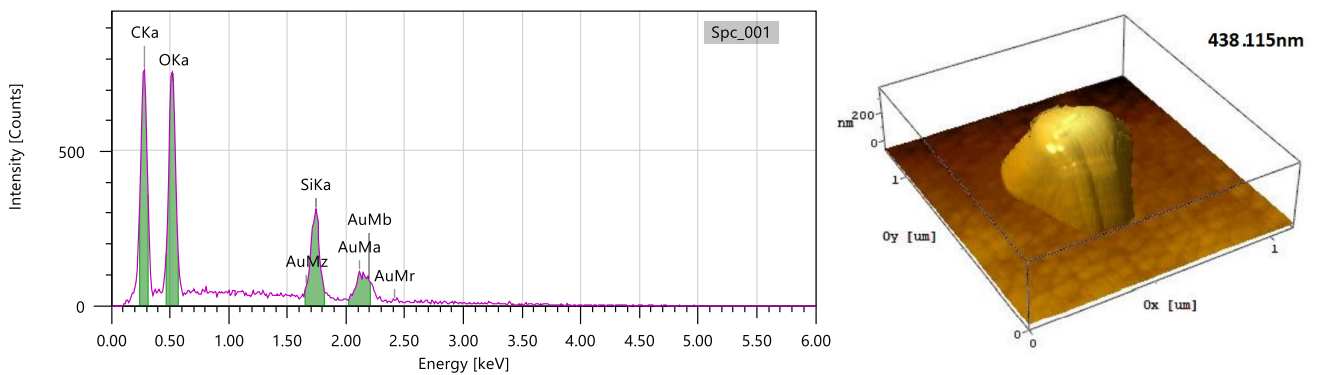

Fig 9. Compositional analysis of structure 2 from **Fig 4** using EDX with AFM view of height

The intent is to build structures with high $SiO_2$ purity, though limited by the high C content and the unavailability to use $H_2$, $O_2$ or $H_2O$ jets to purify the structure *[37], [39]*. *Geier et al (2014) [39]* report the full removal of C content of deposited Pt structures using $H_2O$ vapour jets. As transition element and not a metal, the rate of C desorption from the substrate is reduced with the $H_2O$ vapour/$O_2$ jets compared to other metals, but still presenting a significant reduction and sensitivity to the process *[40]*. An increase in the Si and the height of the structure is observed with dwell time, but well limited by the high C content of the nanostructures; and in comparison line profiles have higher C content due their horizontal growth compared to vertical growth of the pillar profiles. The C content we report from our experimental measurements on our lines and pillar profiles has minimum values of 40at% C with the highest value recorded of 57.43at% C. The second structure (**Fig 9**) has 0 at% Si composition that we assign to the missing of the structure and deformation of the surface, the EDX being able to identify in this case only contamination gasses from air. An average value of the at% composition of Si of 12.08at% is obtained from the 12 structures that present contents of Si higher than 0at% (the one structure without any Si was not taken into account for the average calculation).

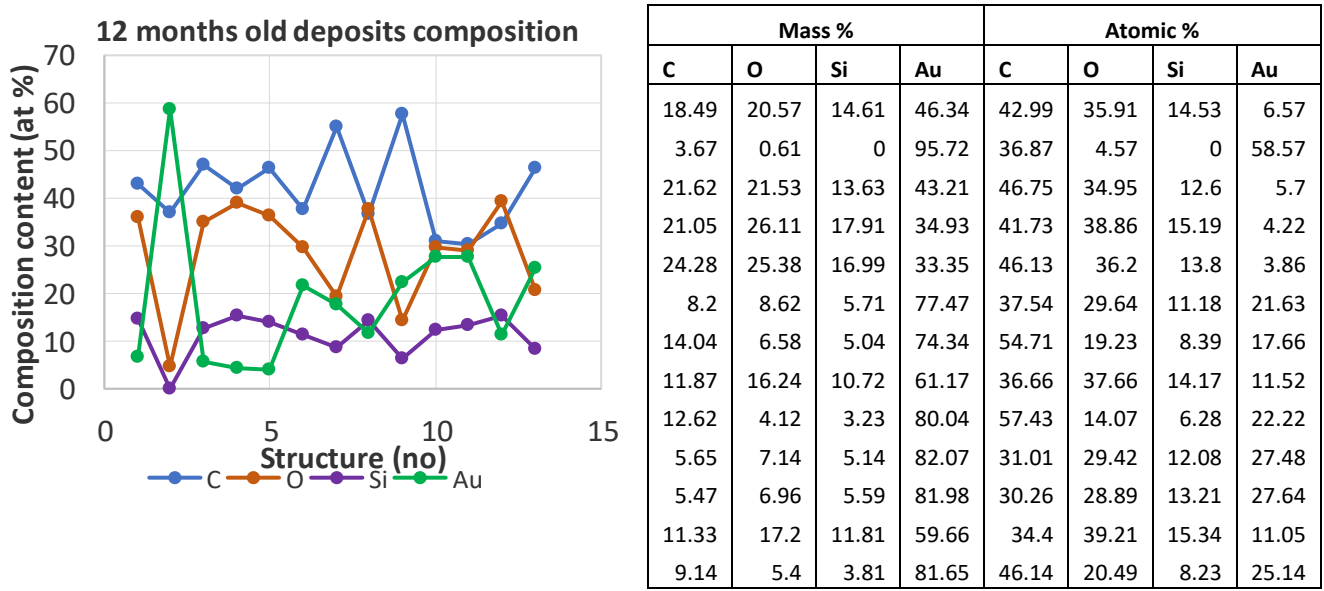

| | Mass % | | | | Atomic % | | | |
|---|---|---|---|---|---|---|---|---|
| | C | O | Si | Au | C | O | Si | Au |
| | 18.49 | 20.57 | 14.61 | 46.34 | 42.99 | 35.91 | 14.53 | 6.57 |
| | 3.67 | 0.61 | 0 | 95.72 | 36.87 | 4.57 | 0 | 58.57 |
| | 21.62 | 21.53 | 13.63 | 43.21 | 46.75 | 34.95 | 12.6 | 5.7 |
| | 21.05 | 26.11 | 17.91 | 34.93 | 41.73 | 38.86 | 15.19 | 4.22 |
| | 24.28 | 25.38 | 16.99 | 33.35 | 46.13 | 36.2 | 13.8 | 3.86 |
| | 8.2 | 8.62 | 5.71 | 77.47 | 37.54 | 29.64 | 11.18 | 21.63 |
| | 14.04 | 6.58 | 5.04 | 74.34 | 54.71 | 19.23 | 8.39 | 17.66 |
| | 11.87 | 16.24 | 10.72 | 61.17 | 36.66 | 37.66 | 14.17 | 11.52 |
| | 12.62 | 4.12 | 3.23 | 80.04 | 57.43 | 14.07 | 6.28 | 22.22 |
| | 5.65 | 7.14 | 5.14 | 82.07 | 31.01 | 29.42 | 12.08 | 27.48 |
| | 5.47 | 6.96 | 5.59 | 81.98 | 30.26 | 28.89 | 13.21 | 27.64 |
| | 11.33 | 17.2 | 11.81 | 59.66 | 34.4 | 39.21 | 15.34 | 11.05 |
| | 9.14 | 5.4 | 3.81 | 81.65 | 46.14 | 20.49 | 8.23 | 25.14 |

Fig 10. Atomic % elemental content of 12 months old structures

The EDX composition analysis reveals the presence of four elements in all structures, Si, O, C and extra N, due to contamination and prolonged exposure to air. In **Fig 11** the 8 months old deposits compositions and 12 hours old deposits compositions are presented. A higher C concentration is observed, with a presence of Si of only 12at% without the addition of other gases during the deposition process. The content of Si increases for the 12 hours old deposits up to values of 16-17at%, but still lower than the reported values of 32at% for a pure $SiO_2$ structure.

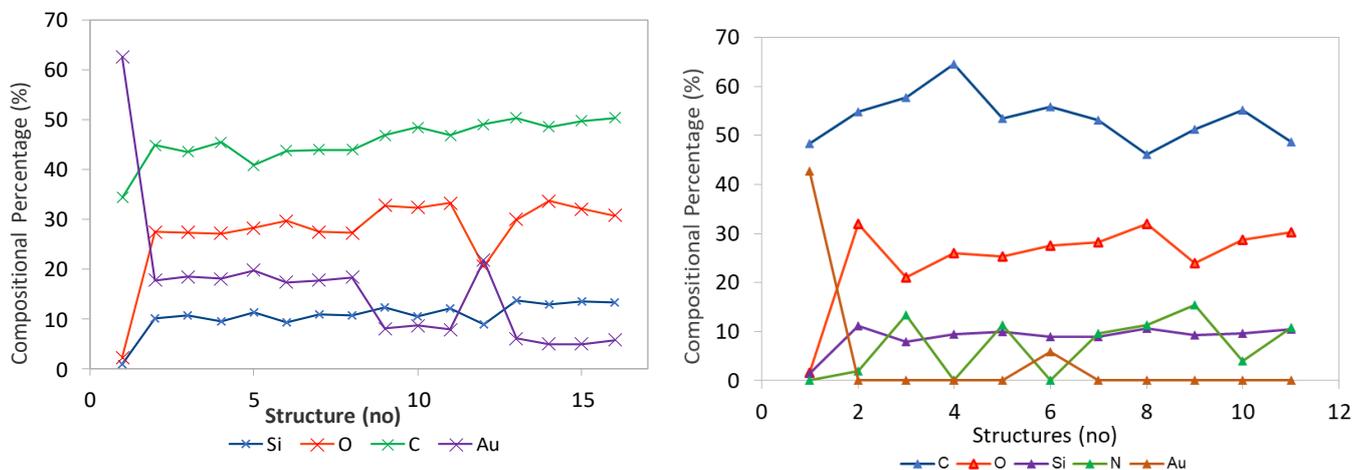

Fig 11. Elemental composition of the $Si(OEt)_4$ deposited nanostructures: 8 months old deposits on irregular Au surface (right) 12 hours old deposits (left)

Structures (in **Fig 11**) of 8 months old deposits have the Au content with values less than 5at%, while the 12 hours old deposits have higher level of Au signal contribution in the compositional analysis of the deposits. The difference between the two measurements suggests the presence of two different surfaces, Au(111)/SiO$_2$, with different thickness of the Au(111) substrate on the SiO$_2$, while the 12 hours substrate is a 100nm Au(1110/SiO$_2$ substrate, the 8 months old substrate is a 200nm Au(111)/SiO$_2$ substrate. The presence of O$_2$ remains constant during time for both the deposited types of structures, though the thickness of the layers is reduced, observed from the high substrate response obtained in the second deposits with only 12 hours wait time (**Fig 11**), the changes in the composition and dimensions of the structures that would be the sign of oxidation and reactivity with the presence of moisture on the substrate from the air, are not observed and are limited in magnitude, without a great impact on the structures over time. Higher C contents are observed for the older structures, which cannot come from the C contained in the Si(OEt)$_4$ at the deposition time, but to the accumulated C during the 8 months the sample was exposed to air.

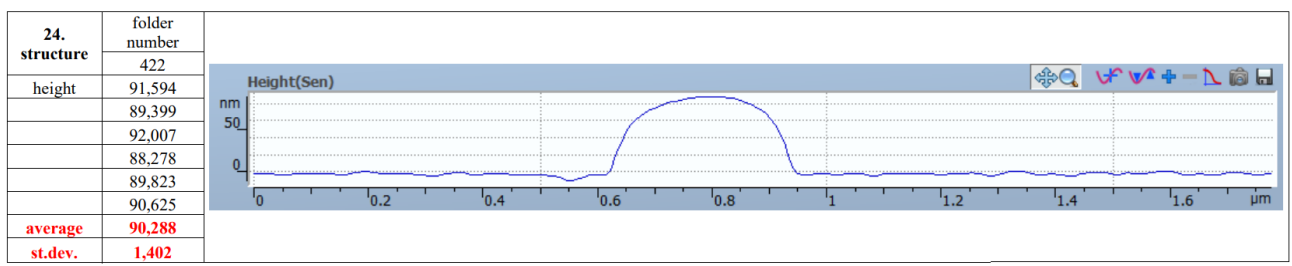

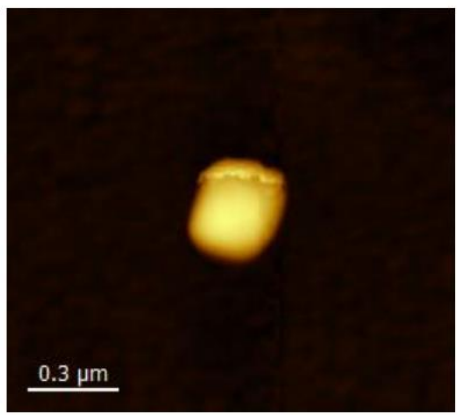
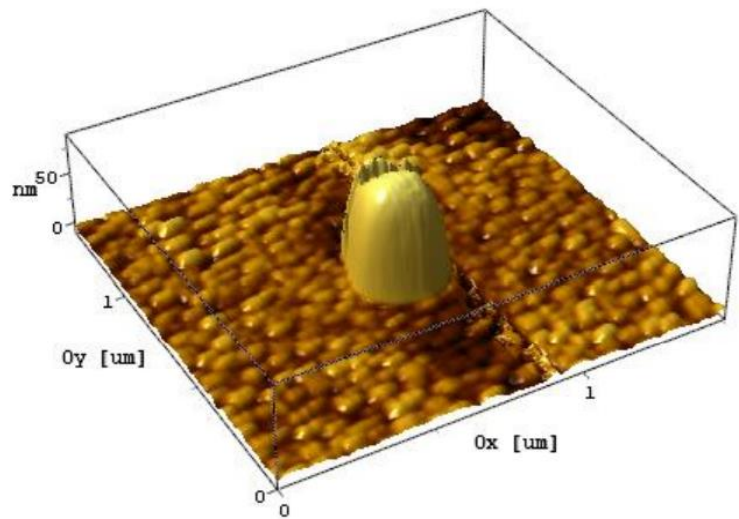

Fig 12. AFM measurement of pillar structure 1 in **Fig 8**. The magnitude of the structural modifications in the pillar can be observed through changes in the height and composition of the nanostructures. For the present structure no oxidation is observed, that would create changes in the morphology, shape and radical changes in height. The height of the structure is 92.007nm at the highest point with an average height of 90.288nm. A substrate with high granulation starts to appear as a result of aging of the SiO$_2$/Au(111)

surface; it is known for the Au(111) substrates to present high Au granulation organized in Chevron patterns on the surface; our Au(111) substrate presents similar behaviour as the one of the substrate reconstruction of Au(111) surface of *Allmers and Donath (2009) [41]*. A lateral width of the deposit of 0.32µm is observed, the growth of the pillar was higher in plane than vertically.

For the 12 hours old structures the C content is under 50at%, while the C content in the case of the 8 months old deposits increases to over 60at%. Another sign of exposure to air of the deposited structure is the presence of small amounts of other atmospheric gasses regularly found in the breathable air in rooms, as $N_2$. The $H_2O$ molecules cannot be determined as water singlets, dimers or trimers using EDX analysis, but an increase in the O content is obtained; our present data shows O content is almost constant in both deposited structures and as well after 12 months from the deposition time. The structures on a new, clean and undamaged Au(111) substrate are unlikely to be affected by $H_2O$ in the air as Si – O bonds are strong bonds with dissociation energies in the range of ~452kJ/mol, while a H – O bond has a value of 467kJ/mol; it is very unlikely for the Si – O bonds to break and form bonds to H.

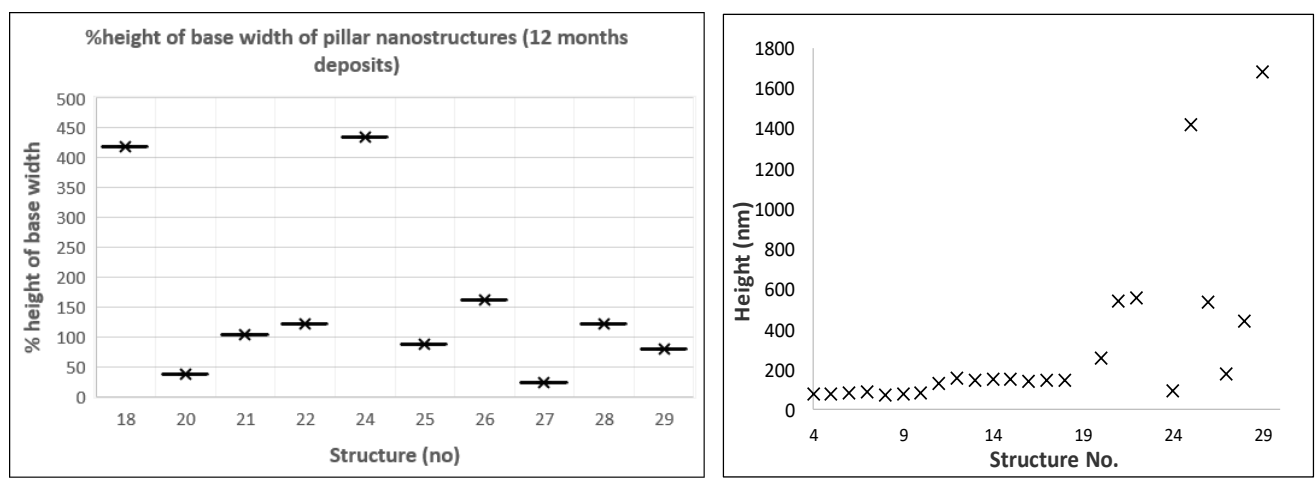

Fig 13. Height of 12 months old structures with the increase of dwell time from 350us to 5030us (right) % height of the base width (left). The width of the structures (see % height of the base width) is linearly proportional with the height of the structures. A good example is structure 29 that has the width of the base close to the value of the height of the structure.

The strength of the $SiO_2$ reduces with the water content and moisture in the atmosphere and is a softer material *[19], [20]* compared to the regular metal nanometre structures, rendering it hard to be grown on the vertical. The extent of the pillar base increases with the increase in the height of the structure, while Au, Pt, Fe and Co compounds hardly increase the base when vertically deposited. With heights between 60nm and 1.7um, the 12 months old structures have conserved up to 80% intact; two of the structures collapsed during the 12 months assigned to a lower width of the base. **Fig 13** presents a comparison of the heights with the base widths of the structure, a thinner base percentage compared with the height creating an unstable structure that due to the softness of the material *[21], [22]* leads to the collapse of the pillar structures. The line profiles have preserved intact up to 100% of all deposited structures. The substrate had no influence on the growth of the nanostructures as it was newly purchased, in very good condition without visible grains and tasation. A different effect (height reduction, structure contamination) is observed for the 8 months old deposits, where for the deposition of the structures an older substrate was used without pre-additional cleaning.

In **Table 4**, structures 20 and 27 are the two structures that tilted reducing the conservation percentage of all the nanostructures to 80%. There is a clear difference in the % height of base width compared to the other structures all having over 60% values. Structure 20 presents a value of 36.52% of the base width value compared to the height of the structure, while structure 27 presents a value of 23.72% of the base width value compared to the height of the structure. Percentages lower than 60% increase the possibility of tilting and collapse of the pillar nanostructures, indicating these structures were created with low dwell time (us), high number of deposition loops (2600) and short stand-by time (see **Table 2** for values). Higher dwell time with the same number of loops increase the % height of base width to over 120%, while lowering the loops (1000) and increasing the dwell time (µs) would create very wide bases of the nanopillars with values of the % height of base width of over ~400%.

| Structure (no) | Max height (nm) | Base Width(nm) | %Height of base width |
|---|---|---|---|
| 18 | 155.176 | 650 | 418.88 |
| 20 | 840 | 306.749 | 36.52 |
| 21 | 544.926 | 560 | 102.77 |
| 22 | 594.941 | 720 | 121.02 |
| 24 | 92.007 | 400 | 434.75 |
| 25 | 1439.5 | 1260 | 87.53 |
| 26 | 583.737 | 950 | 162.74 |
| 27 | 1390 | 329.679 | 23.72 |
| 28 | 438.115 | 530 | 120.97 |
| 29 | 1720.31 | 1360 | 79.06 |

Table 4. 12 months old deposits % height of the base width of pillar structures

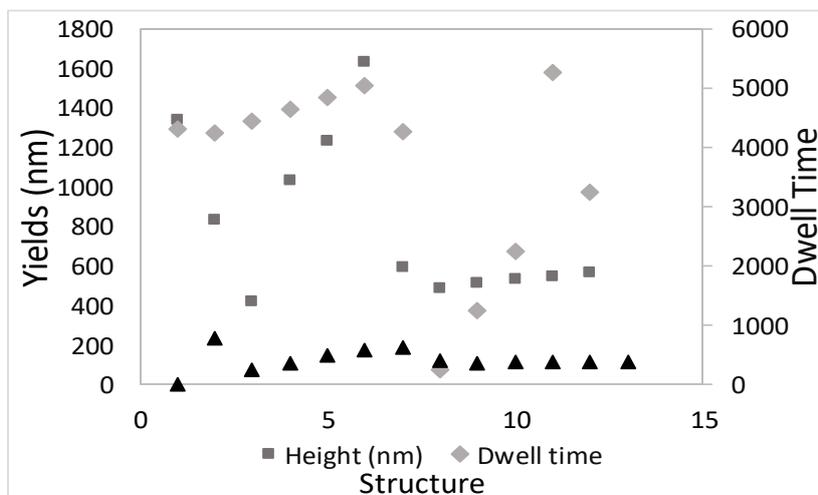

Fig 14. Height/width with dwell time of 8 months old nanopillar structures; (peak values presented in supplementary material). The decrease in height with the increase in the dwell time observed at 750µs and

1200μs compared to the height regime observed at 350μs and 500μs where with the increase in time an increase in the structure height is observed corresponding to a beam limited growth regime *[8], [10], [11]*. For an average 1keV beam current, the dwell time limit to suppress the growth is set to around 700μs. Applied in biomedical and biosensing applications, the $SiO_2$-frameworks enhanced with organic material for mineralization of bone tissues and DNA *[46]* deposited at the nanoscale, and their growth and mineralization, make the focus of highly increasing in importance studies on inhomogeneous (ex. Au(111) substrate with defects) and homogenous (perfect hcp lattice) substrate depositions. Defects in the Au(111) surface are filled with $SiO_2C_x$ material through C bonds between the C in the defects and the remaining C of the deposited molecules. Rolling of already formed $SiO_2$ can take place with higher energy consume through vaporization and electron excitation of the formed ions. The bonding with O-layers at the surface is highly improbable *[48]* as the Au(111) surface would need to undergo an oxidation process, while the most probable process is the integration of Si atoms in the substrate at the deposition moment.

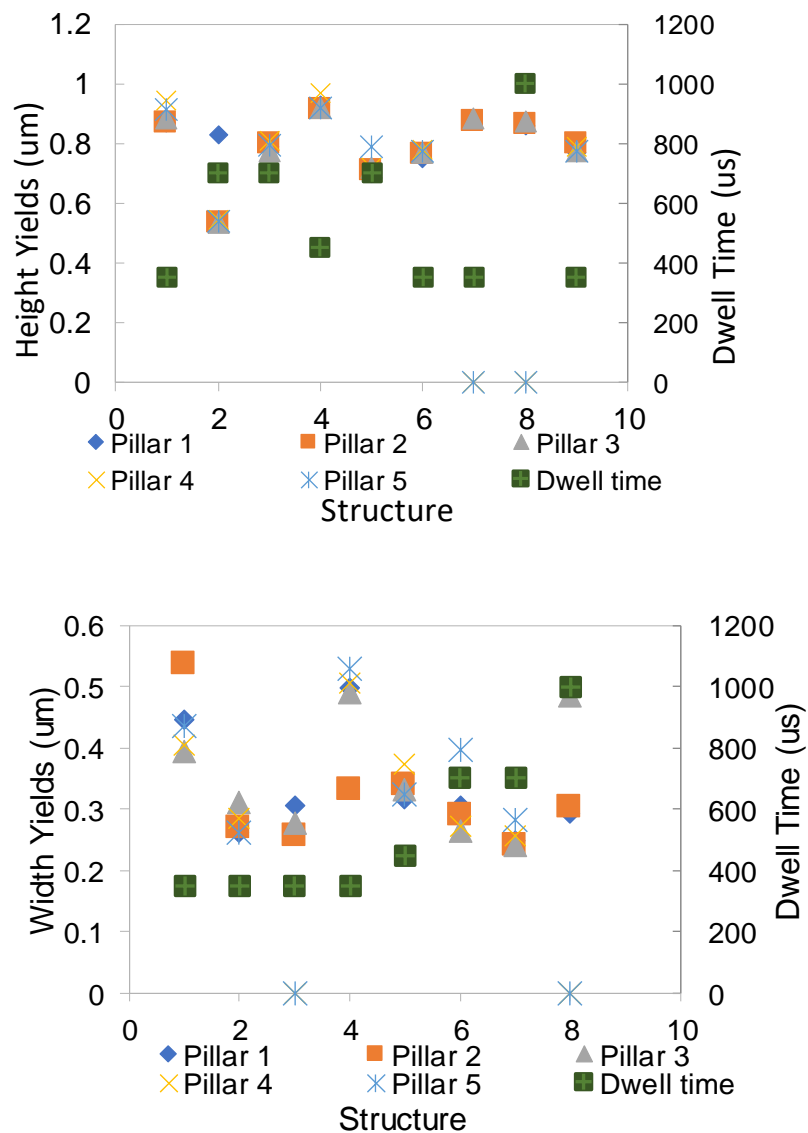

Fig 15. Height/width with dwell time of 12 hours old structures; (a) height (μm) (b) width (μm). The same behaviour as seen in **Fig 14** is observed for the 12 hours old deposits with even more pronounced

differences (**Fig 14**). The behaviour does not change with the analysis of the width increase and decrease, the lateral growth follows the same path, and it is limited by the dwell time and the beam current magnitude. With a Si : O ratio content of 11 : 28, the 12 months old structure exhibit slightly less height and width, still having a high amount of C content, in a process of deposition that did not use additives for purification of deposits. No further purification step was done for the nanostructures using either $O_2$ or $H_2O$ jets. A common behaviour is the increase in the C content of the structures following exposure to air and room temperature. *Sánchez et al (2002) [47]* analyse the process of deposition using electron beam and ion beam depositions with higher Si : O (33 : 15) ratios with the addition of $H_2O$ in the deposition process. *Plank et al (2020) [45]* comparing their deposition studies to the ones of *Sánchez et al (2002) [47]* declare structures of less than 10nm deposited in thin layers (used for coatings in the manufacturing industry) and with composition close to the composition of pure $SiO_2$. A higher O : Si ratio (**Fig 11**) is observed in good agreement with *[45]* and *[47]* for the 12 hours old structures with an average of 29 : 11.

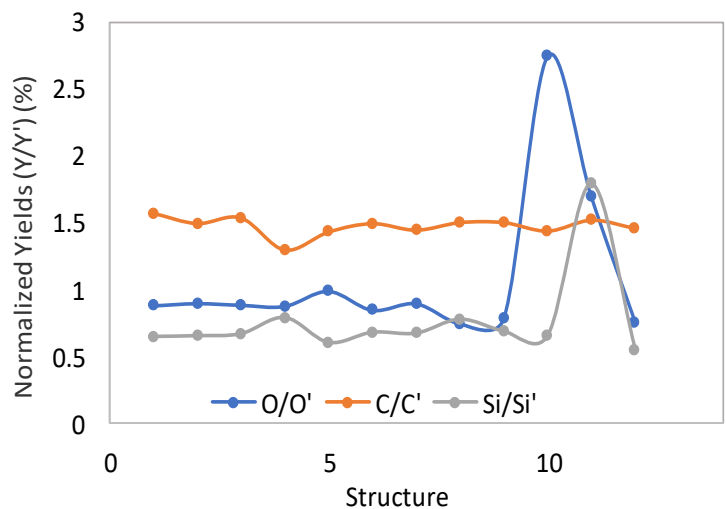

Fig 14. Ratios of C, O and Si for 12 months old deposits (') and 12 hours old deposits ()

For values under 1, the ratios of the 12 months old deposits compared with the 12 hours old deposits show an increase in the compositional content of the structure with the measured element. The C/C' presents the expected behaviour; with time the carbon content of the structures increases, but when it comes to the O and Si, almost constant behaviour is observed in the at% composition of the structures, caused by a phenomenon of evaporation or inoculation of O and Si in the presence of $H_2O$ and N from the air.

**Conclusions**

The behaviour of the $Si(OEt)_4$ at FEBID deposition was repeatedly studied by scientist in different conditions as the precursor is one of the widest used precursors for the deposition of Si and $SiO_2$. The accumulation of moisture through the high number of O atoms in the deposits, creates problems at the deposition of pillar shape structures where the base of the structure is much smaller than the height of the structure. The present study wants to bring up and discuss the different conditions for the deposition of $Si(OEt)_4$, from the point of view of the deposition temperature to the time and exposure to the electron beams and especially air. Results have been obtained for structures between 9nm and up to 0.5um with success, though the deposition of very small structures less than 6nm from Si(OEt)$_4$ still remains a quest.

**Acknowledgements.** We want to acknowledge MP receiving funding from the European Union's Horizon 2020 research and innovation program under the Marie Skłodowska-Curie grant agreement No 722149, and the work of our partner institutions Laboratory of Materials Science, Institute for Nuclear Research, Debrecen, Hungary.

No conflict of interests has been declared.

# Supporting Information for Structural Analysis of Si(OEt)$_4$ Deposits on Au(111)/SiO$_2$ Substrates at Nanometer Scale using Focused Electron Beam Induced Deposition

[1,2]Po-Shi Yuan, [3]Maria Pintea, [3]Nigel Mason

[1]Zeiss Gmbh, Rosdorf, Germany

[2]University of Darmstadt, Darmstadt, Germany

[3]University of Kent, School of Physical Sciences, Canterbury, CT2 7NH, UK

Table 1: 28pA, 1keV beam characteristics deposition

| Structure | Height (nm) | Width (nm) | Deposition time (s) | Dwell time | Loops |
|---|---|---|---|---|---|
| Profile 1 | 123.1 | 350 | 182.5 | 350 | 1000 |
| Profile 7 | 128 | 350 | 182.5 | 350 | 1000 |
| Profile 2 | 129.3 | 350 | 182.5 | 350 | 1000 |
| Profile 6 | 126.5 | 350 | 182.5 | 350 | 1000 |
| Profile 3 | 168.8 | 350 | 182.5 | 350 | 1000 |
| Profile 4 | 160.8 | 350 | 182.5 | 350 | 1000 |
| Profile 5 | 165.1 | 350 | 182.5 | 350 | 1000 |
| Profile 8 | 88.9 | 341 | 370.31 | 700 | 1300 |
| Profile 14 | 89.9 | 341 | 370.31 | 700 | 1300 |
| Profile 9 | 87 | 320 | 351.2 | 700 | 1300 |
| Profile 13 | 82 | 320 | 351.2 | 700 | 1300 |
| Profile 10 | 62.4 | 310 | 342.1 | 700 | 1300 |
| Profile 11 | 62.9 | 340 | 369.4 | 700 | 1300 |
| Profile 12 | 62.4 | 330 | 360.3 | 700 | 1300 |

|  | Line1 (nm) | Line 2 (nm) | Line 3 (nm) | Line 4 (nm) | Line 5 (nm) | Dwell Time (us) | Loops (no) |
|---|---|---|---|---|---|---|---|
| D1 Height | 163.8 | 168.8 | 140.8 | 177 | 152.8 | 700 | 700 |
| D1 Weight | 370 | 300 | 280 | 300 | 330 | 700 | 700 |
| **Dep time** | **241.3** | **207** | **197.2** | **207** | **221.7** | | |
| D2 Height | 311.6 | 296.2 | 283.9 | 287.1 | 291.7 | 450 | 500 |
| D2 Weight | 330 | 330 | 330 | 410 | 330 | 450 | 500 |
| **Dep time** | **134.25** | **134.25** | **134.25** | **152.25** | **134.25** | | |
| D3 Height | 103.2 | 107.1 | 117.1 | 120.4 | 116.6 | 700 | 500 |
| D3 Weight | 320 | 280 | 280 | 370 | 300 | 700 | 500 |
| **Dep time** | **172** | **158** | **158** | **189.5** | **165** | | |
| D4 Height | 146.2 | 196 | 161.6 | 173.4 | 175.7 | 350 | 900 |
| D4 Weight | 240 | 320 | 350 | 280 | 280 | 350 | 900 |
| **Dep time** | **135.6** | **160.8** | **170.25** | **148.2** | **148.2** | | |
| D5 Height | 289.6 | 288 | 290.7 | 0 | 0 | 350 | 1300 |
| D5 Weight | 272 | 247 | 239 | 0 | 0 | 350 | 1300 |
| **Dep time** | **183.76** | **172.385** | **168.745** | | | | |
| D6 Height | 278.6 | 283.6 | 291.9 | 0 | 0 | 1000 | 1300 |
| D6 Weight | 285 | 291 | 309 | 0 | 0 | 1000 | 1300 |
| **Dep time** | **430.5** | **438.3** | **461.7** | | | | |
| D7 Height | 158.5 | 168.5 | 176.3 | 188.6 | 173.3 | 350 | 1300 |
| D7 Weight | 322 | 289 | 289 | 289 | 330 | 350 | 1300 |
| **Dep time** | **206.51** | **191.495** | **191.495** | **191.495** | **210.15** | | |
| D8 Height | 329.8 | 346.1 | 363.8 | 401 | 379.2 | 700 | 1300 |
| D8 Weight | 445 | 520 | 396 | 396 | 384 | 700 | 1300 |
| **Dep time** | **464.95** | **533.2** | **420.36** | **420.36** | **409.44** | | |
| D9 Height | 237.9 | 224.4 | 203.6 | 259.5 | 199.4 | 350 | 900 |
| D9 Weight | 361 | 299 | 289 | 330 | 309 | 350 | 900 |
| **Dep time** | **173.715** | **154.185** | **151.035** | **163.95** | **157.335** | | |

Table 2: 24pA, 1keV beam characteristics deposition

Additional structural compositional data is presented in Table 3 and Fig 1, 2, 3.

Table 3: Ratios of the compositional elements

|  | Dwell time(us) | Loops | C/C' | O/O' | Si/Si' |
|---|---|---|---|---|---|
| Background | 0 | 0 | 0.709 | 1.438 | 0.643 |
| Profile 1 | 350 | 1000 | 0.816 | 0.856 | 0.91 |
| Profile 2 | 350 | 1000 | 0.754 | 1.3 | 1.35 |
| Profile 3 | 350 | 1000 | 0.703 | 1.042 | 1.011 |
| Profile 4 | 350 | 1000 | 0.764 | 1.115 | 1.152 |
| Profile 5 | 350 | 1000 | 0.782 | 1.076 | 1.044 |
| Profile 8 | 700 | 1300 | 0.88 | 1.16 | 1.382 |
| Profile 9 | 700 | 1300 | 1.05 | 1.013 | 1 |
| Profile 10 | 700 | 1300 | 0.914 | 1.389 | 1.301 |
| Profile 11 | 700 | 1300 | 0.858 | 1.128 | 1.281 |
| Profile 12 | 700 | 1300 | 0.967 | 1.076 | 1.192 |

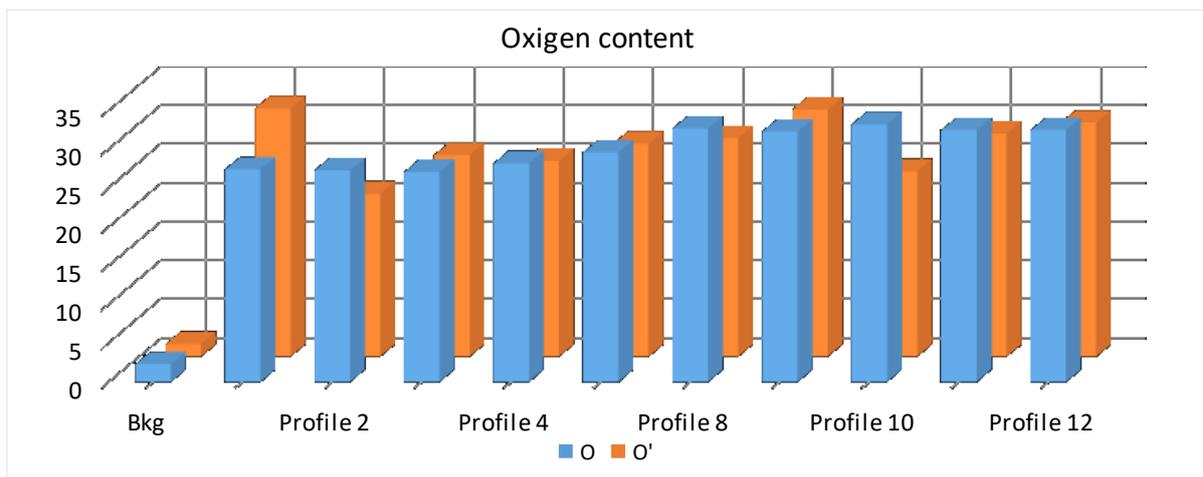

Fig 1. Oxygen content of 12 hours old deposits () and 8 months old deposits (')

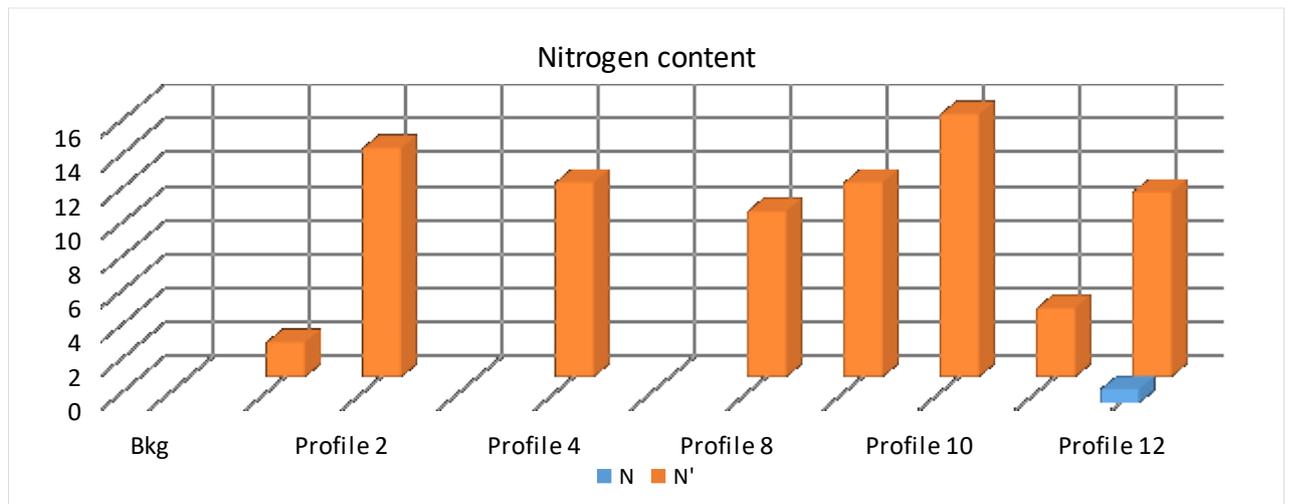

Fig 2. Nitrogen content of 12 hours old deposits () and 8 months old deposits (')

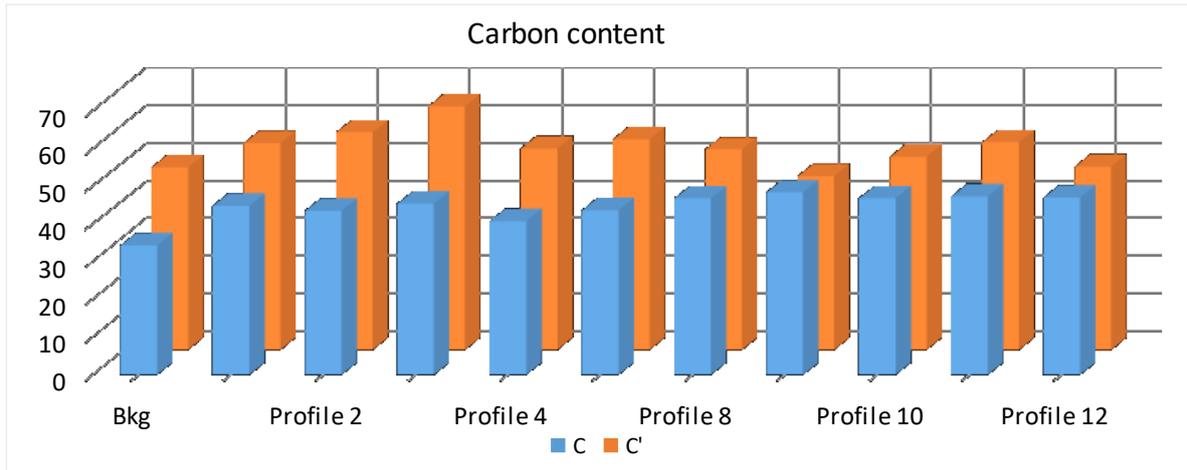

Fig 3. Carbon content of 12 hours old deposits () and 8 months old deposits (')